# Reconfigurable Power Electronics Topologies

Haochen Li, Jonathan Shek

Institute for Energy Systems, School of Engineering, The University of Edinburgh, The King's Buildings, Mayfield Road, Edinburgh, EH9 3DW, United Kingdom

***Abstract*** This paper presents two novel topologies for automatically transforming power converter topology from three-phase 3-level cascaded H-bridge to three-phase 2-level converter design. These techniques are implemented by flicking specific switches to rearrange circuit connections. The switches can be controlled by signals in order to realize automation.

## 1. Introduction

In this paper, topologies are designed so that they can be transformed from three-phase 3-level converters to 2-level according to demands. The processes and transformations are simulated with MATLAB/SIMULINK. The topologies have 6 separate DC sources as inputs and produce three-phase 3-level or 2-level AC outputs, with signal generators including step signal source, pulse generator and 2-level PWM generator.

Reconfigurable power converter topologies have potential to be used in HVDC transmission. As the HVDC is commonly used to transmit power nowadays and power converter are essential to transfer power from DC sources into three-phase AC voltage in order to be used for grid.

The designed topologies also have a wide application potential to build power converters for renewable energy. For example, one of the most popular renewable energy nowadays – wind power needs this technique basically. Cascaded inverters are essential to compose the joint between AC grid and separate DC sources which is usually used as renewable energy sources such as wind turbines, photovoltaics or fuel cells [1].

Also reconfigurable power converter provides mobility and flexibility in practice. The circuits build by it can be transformed between 3-level AC output and 2-level AC output easily by controlling signal. It inspires a method to build reconfigurable topologies comprised by different kinds of multilevel converter and 2-level converter. And the output AC voltage will change after transformation, which is a potential application for different requirement of voltage.

MATLAB/SIMULINK software is generally used to simulate dynamic systems with a block diagram environment. It is integrated in the MATLAB and assists design, simulation, test, analysis and verification of embedded systems [2]. This software is used in this work to design, test and analyse the reconfigurable power converter topologies.

## 2. Background

Three-phase voltage source inverters are essential in medium and high level power applications to supply a three-phase voltage source to AC grid. Its amplitude, phase and frequency can be changed as required [3].

The concepts of multilevel power converters appeared in 1975 [4] and have developed a lot since then. The multilevel converter synthesizes a sinusoidal voltage with several voltage levels that obtained from capacitor voltage sources [5]. Multilevel converter has several advantages compared to a traditional 2-level converter including better staircase waveform quality, smaller common-mode (CM) voltage, drawing input current with less distortion and being able to operate at various frequencies. However, a great number of power semiconductors switches are needed and causes the overall system to be expensive as well as complicated [1].

There are three kinds of multilevel converter structure majorly used in industry: cascaded H-bridge converter with separate DC sources, diode clamped (neutral clamped) and flying capacitors (capacitor clamped) [1].

In this paper, 3-level cascaded H-bridge power converter will be used to design three-phase reconfigurable power converter topologies, which provides foundation and inspiration for further reconfigurable topologies using other kinds of multilevel power converter structures or power converter with higher level.

## 3. Design of Reconfigurable Topologies

### 3.1 Explanation

To build a reconfigurable power converter topology, it is necessary to combine the two topologies – three-phase 3-level cascaded H-bridge converter and three-phase 2-level bridge inverter. At first the two topologies need to be simulated and tested to ensure that they are functional. Then an analysis will be performed to find the way combining them, and the least amount of switches shall be used while all functions work naturally.

### 3.2 Construction of three-phase 3-level cascaded H-bridge converter

The three-phase 3-level cascaded H-bridge converter is designed and tested first, whose block diagram is shown in Fig.5, and the output waveform in Fig.6.

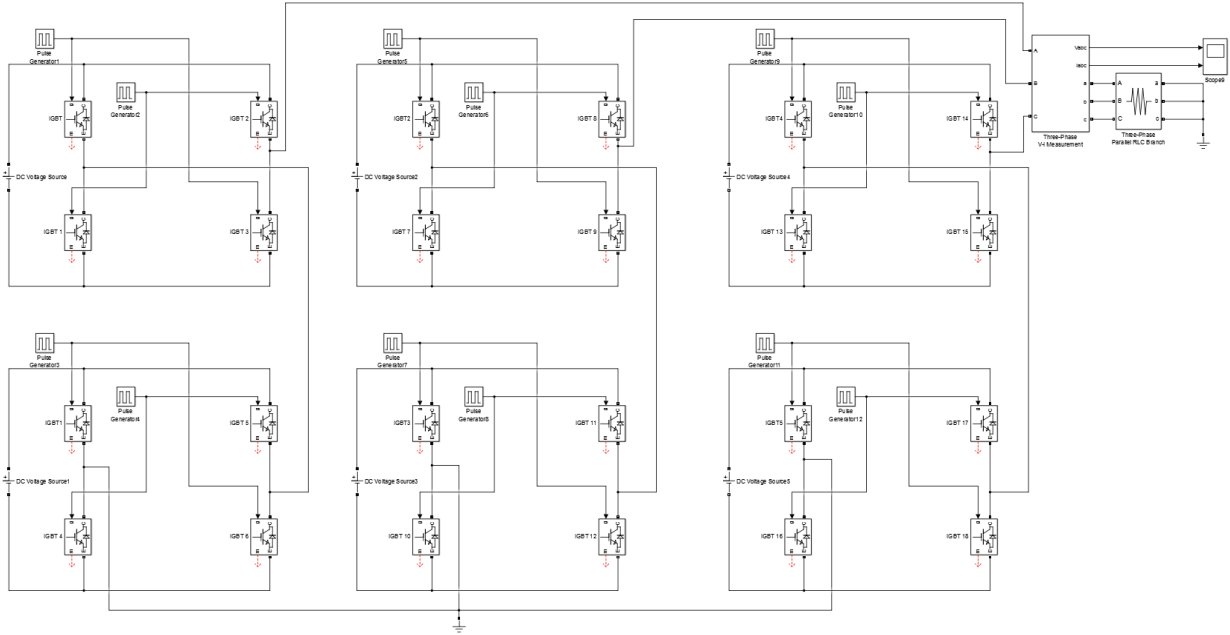

**Fig.5 Block diagram of the three-phase 3-level cascaded H-bridge converter**

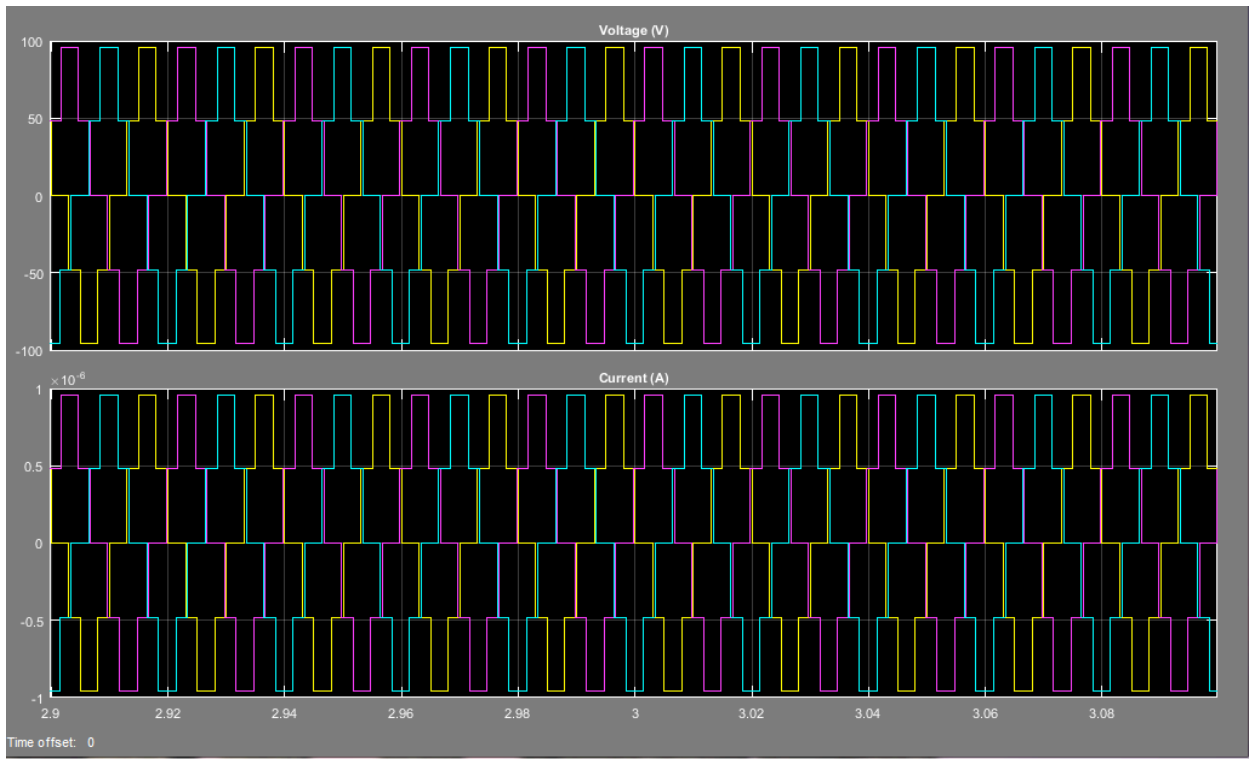

Fig.6 Output waveform of the three-phase 3-level cascaded H-bridge converter

## 3.3 Construction of three-phase 2-level bridge inverter

Next, the three-phase 2-level bridge inverter is designed and tested, whose block diagram is shown in Fig.7, and the output waveform in Fig.8. Note that the state of the two ideal switches are 0, representing that they are open.

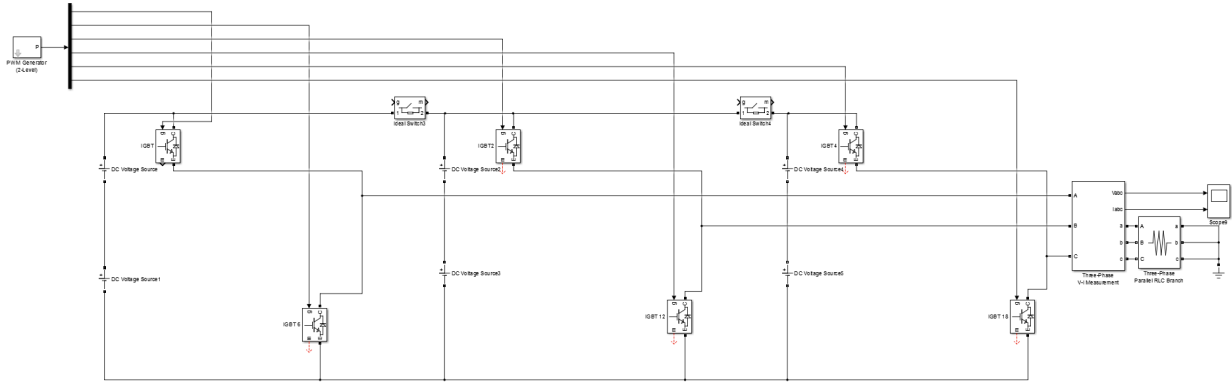

Fig.7 Block diagram of the three-phase 2-level bridge inverter

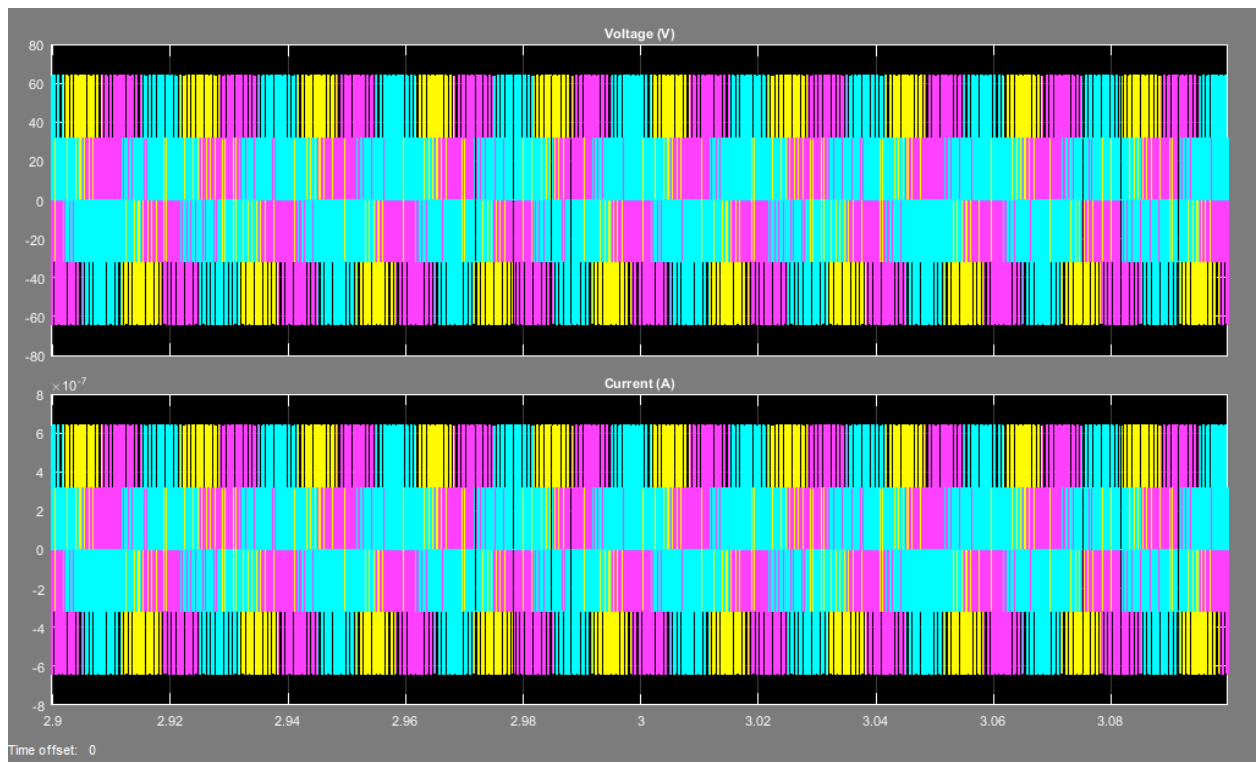

Fig.8 Output waveform of the three-phase 2-level bridge inverter

## 3.4 Combination of the two topologies into a reconfigurable topology

As both topologies (Fig.5, Fig.7) run well, they can be combined together and proceed to the posterior process. The block diagram of the overall reconfigurable topology is shown in Fig.9, which has output waveforms as Fig.10 and Fig.11. Here we use value 1e12 as the resistance of the ideal switches, and they are driven by 2 step signals with step time of 3s, in other words, the topology is transformed with a time interval of 3s. Also each DC voltage sources is set to output 48V.

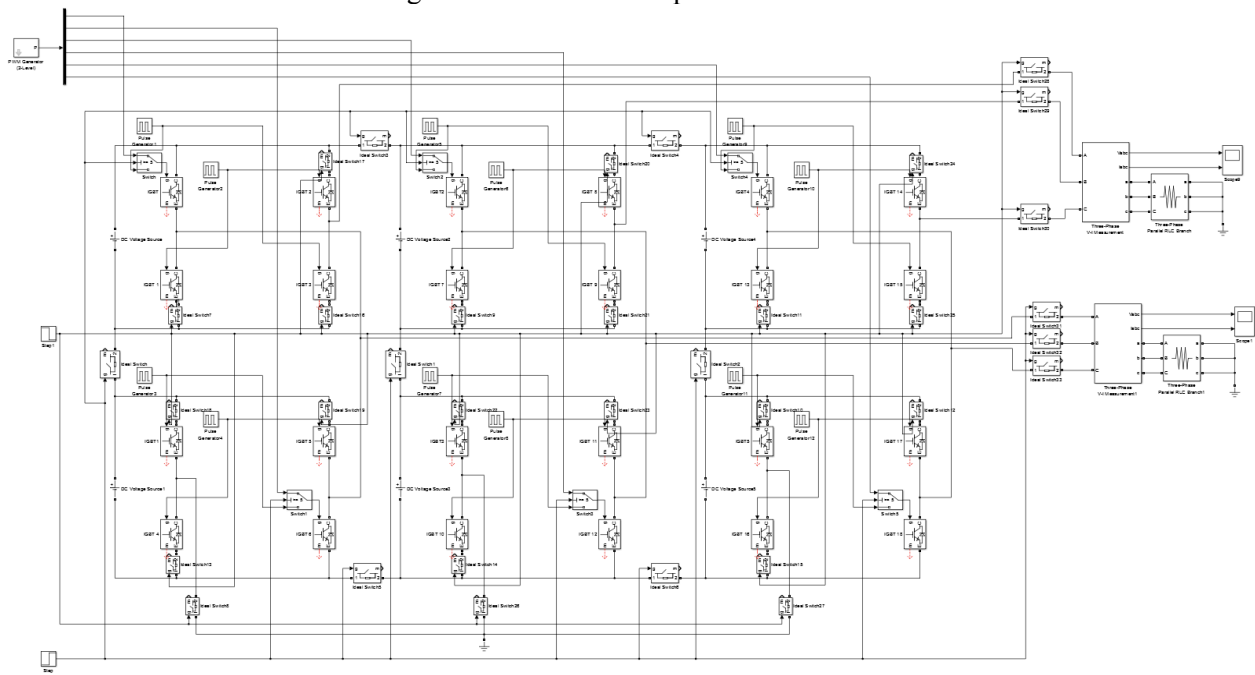

Fig.9 Block diagram of reconfigurable topology 1

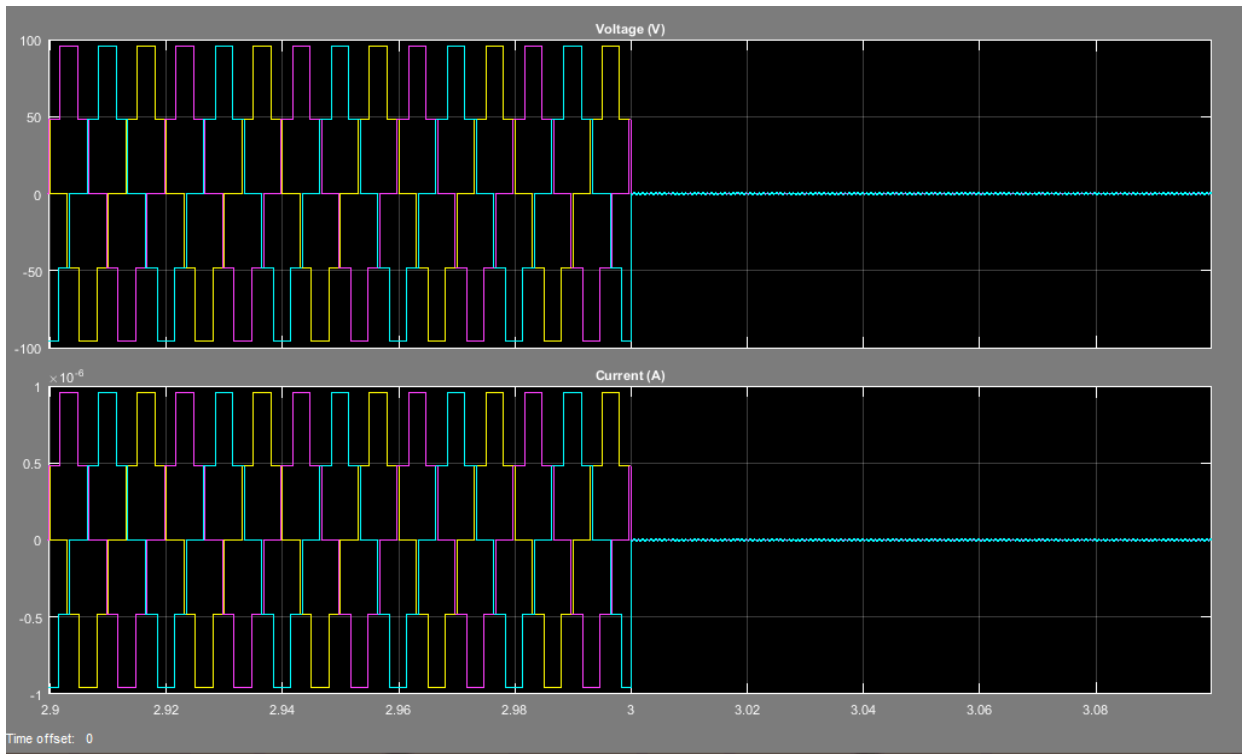

**Fig.10 Output waveform of three-phase 3-level cascaded H-bridge inverter among reconfigurable topology 1**

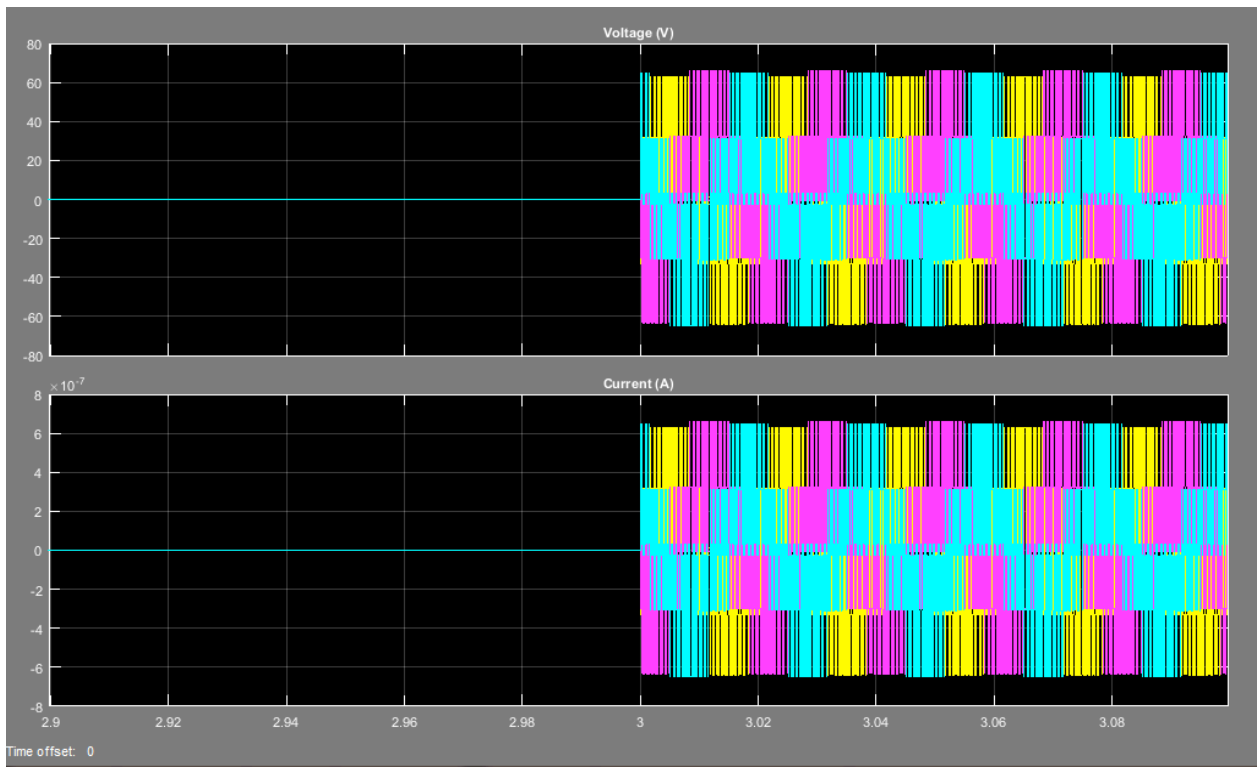

**Fig.11 Output waveform of three-phase 2-level bridge inverter among reconfigurable topology 1**

### 3.5 Extension of three-phase 2-level bridge inverter

Also there is an extension topology from three-phase 2-level bridge inverter (Fig.7), which use H-bridge IGBTs instead of separated pulse generators. This topology is designed and tested, the block diagram of which is shown in Fig.12, and the output waveform in Fig.13. As shown in Fig.12, it is neater than before (Fig.9).

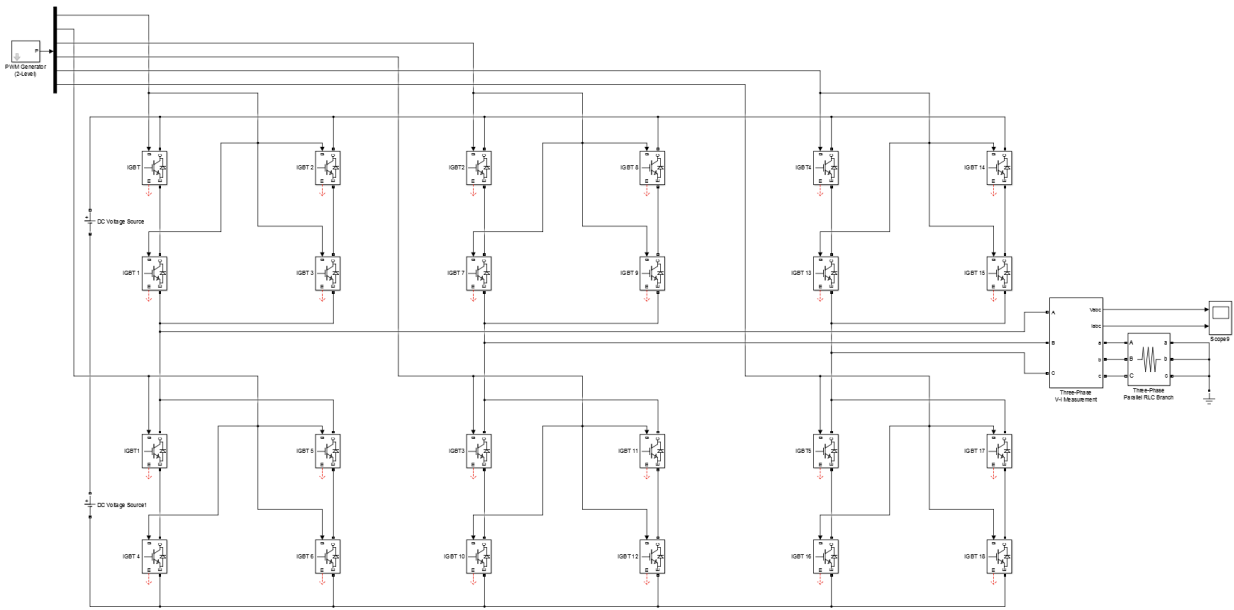

**Fig.12 Block diagram of the extension three-phase 2-level bridge inverter**

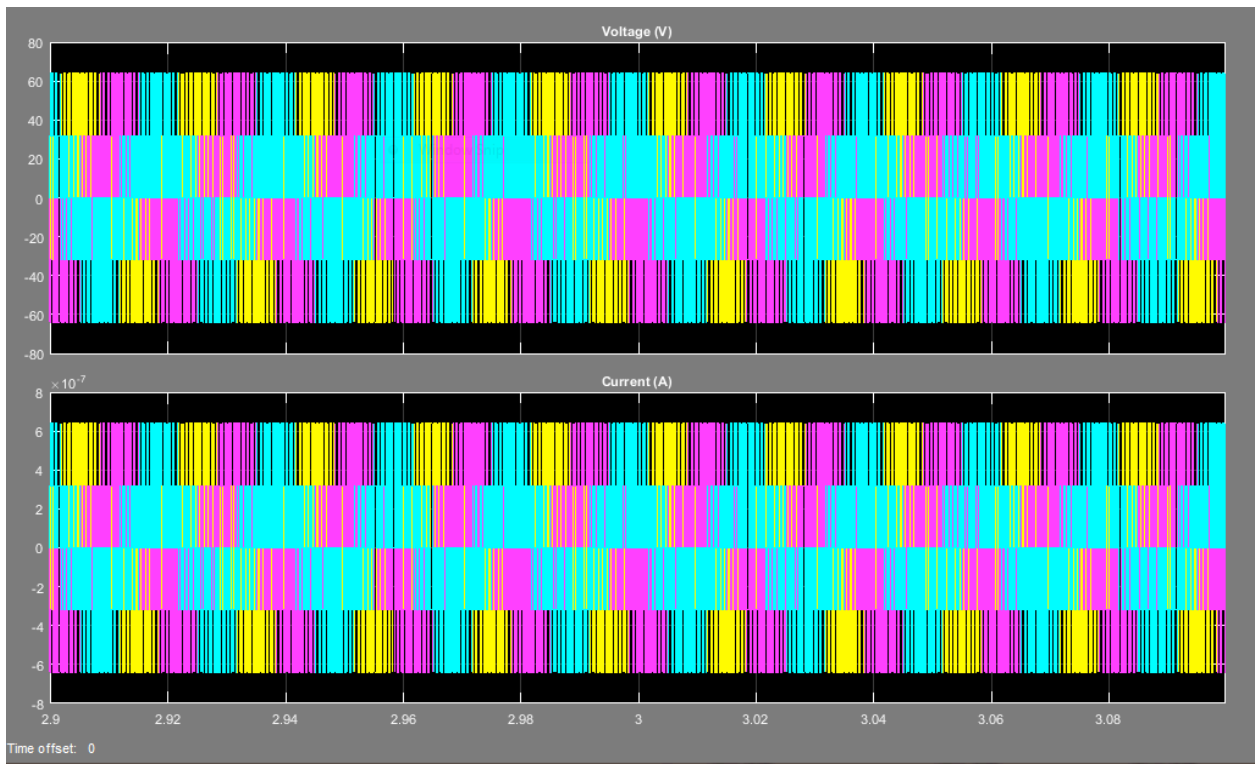

**Fig.13 Output waveform of the extension three-phase 2-level bridge inverter**

## 3.6 Another reconfigurable topology

Another proposed reconfigurable topology is composed by three-phase 3-level H-bridge converter (Fig.5) and the extense three-phase 2-level bridge inverter (Fig.12). The block diagram of overall reconfigurable topology is shown in Fig.14 and has output waveforms as Fig.15 and Fig.16.

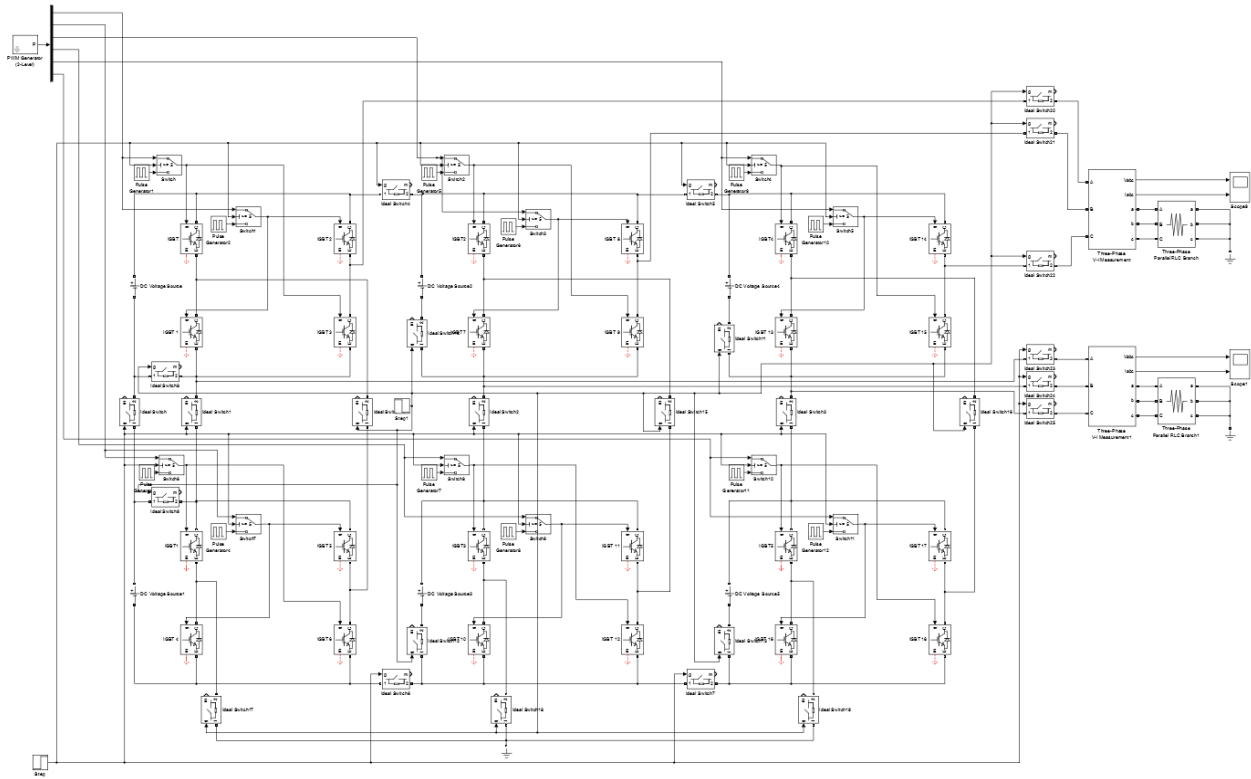

**Fig.14 Block diagram of reconfigurable topology 2**

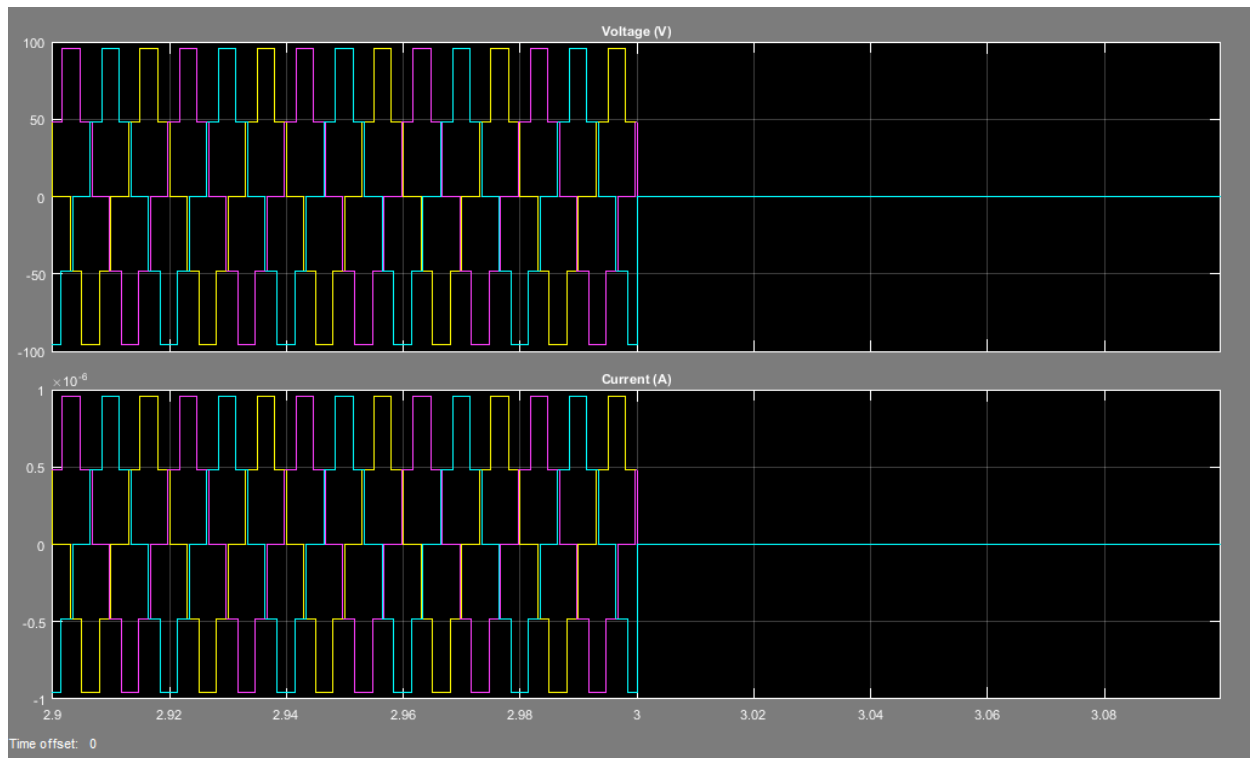

**Fig.15 Output waveform of three-phase 3-level cascaded H-bridge inverter among reconfigurable topology 2**

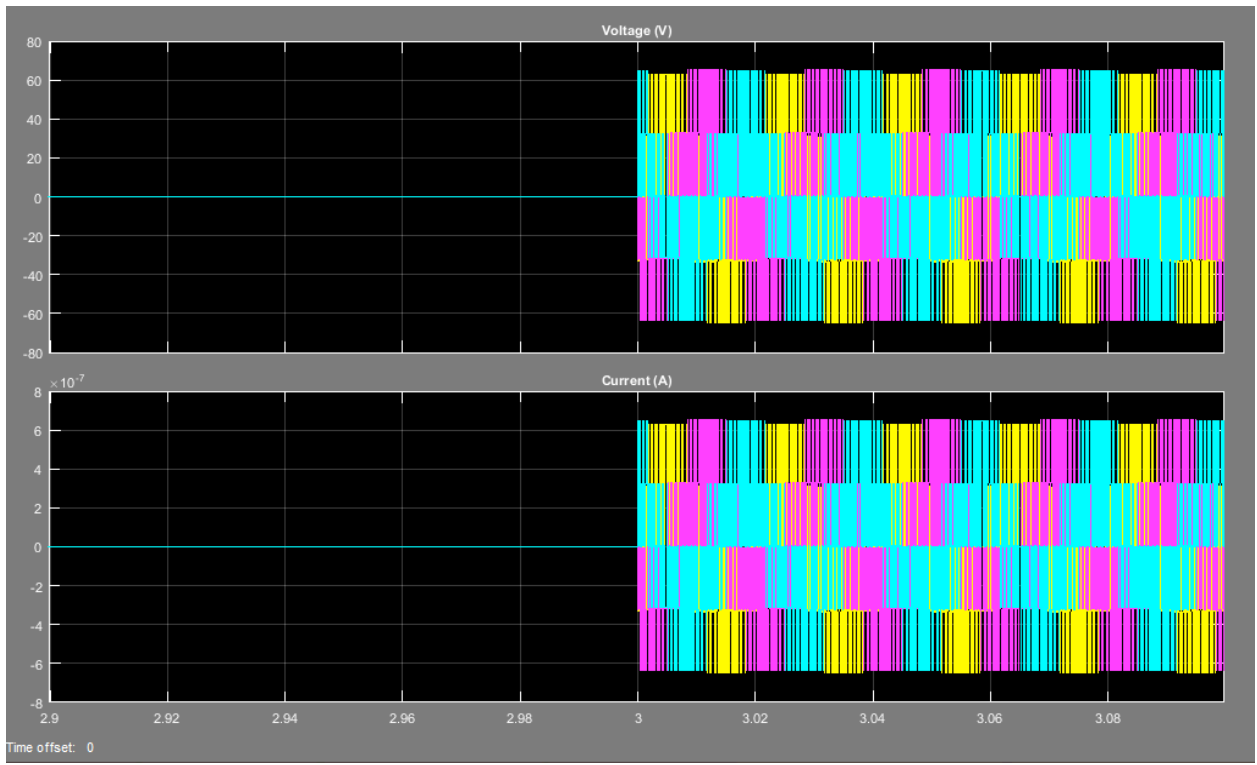

**Fig.16 Output waveform of extension three-phase 2-level bridge inverter among reconfigurable topology 2**

## 4. Features and Analysis

### 4.1 The amount of used switches

The total number of different switches are counted and shown in Tab.1.

It is apparent that the total amounts of switches are about the same, while the second topology uses slight fewer switches.

|              | Reconfigurable topology 1 | Reconfigurable topology 2 |
|--------------|---------------------------|---------------------------|
| IGBT         | 24                        | 24                        |
| 3-way switch | 6                         | 12                        |
| Ideal switch | 34                        | 26                        |
| Total        | 64                        | 62                        |

**Tab.1 Statistics of switches**

### 4.2 Phase voltage difference range

The three-phase 3-level cascaded H-bridge converters act ideally in both reconfigurable topologies as all the phase voltage are the same and equal to 96V which equals to the total voltage of DC voltage sources in series. However, as we can see in Fig.17 and Fig.18, the phase voltage of both 2-level bridge inverter are different in amplitude, while that of reconfigurable topology 1 fluctuate more acutely as the maximum difference range of reconfigurable topology 1 is 3V compared to 1V of reconfigurable topology 2. Thus, reconfigurable topology 2 has more stable output voltage and works better.

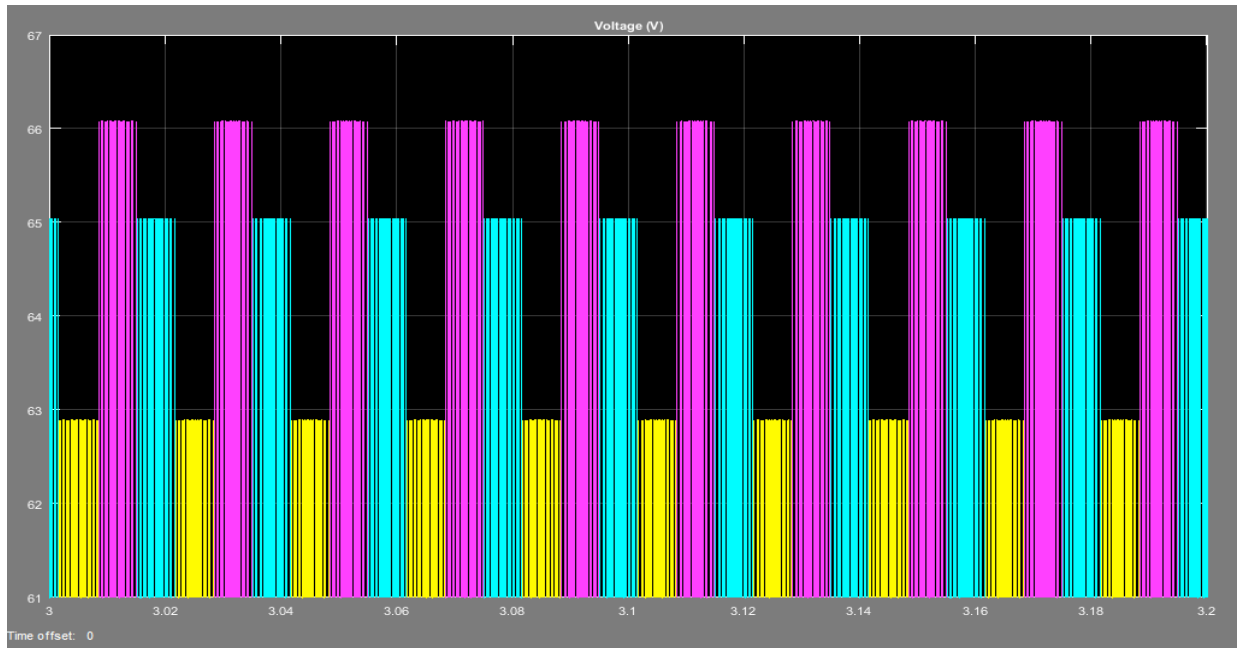

**Fig.17 Illustration on phase voltage difference of 2-level bridge inverter in reconfigurable topology 1**

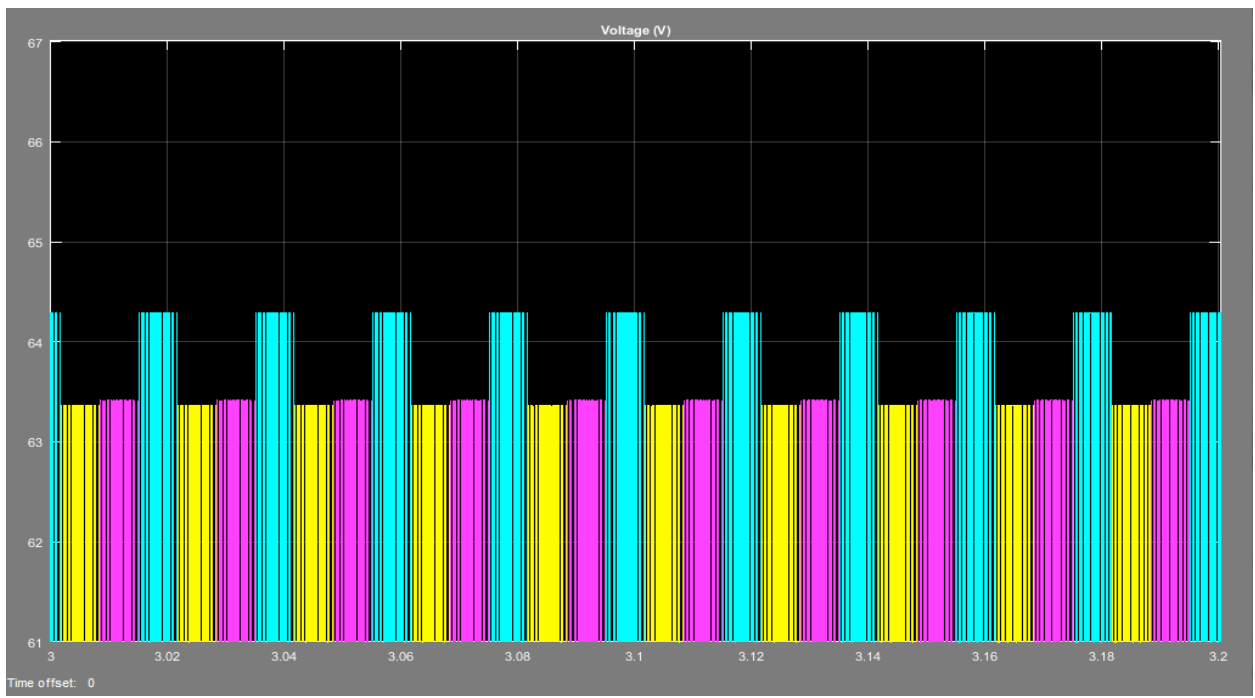

**Fig.18 Illustration on phase voltage difference of 2-level bridge inverter in reconfigurable topology 1**

## 4.3 Effect of ideal switches resistance

During experiments it is found that the resistance of ideal switches influences the performance of reconfigurable topologies significantly. If the resistance of all ideal switches in reconfigurable topology 1 are changed from 1e12 to 1e9, we can get output waveform shown in Fig.19 and Fig.20 (compared to Fig.10 and Fig.11). And reconfigurable topology 2 results in outputs as Fig.21 and Fig.22 (compared to Fig.15 and Fig.16).

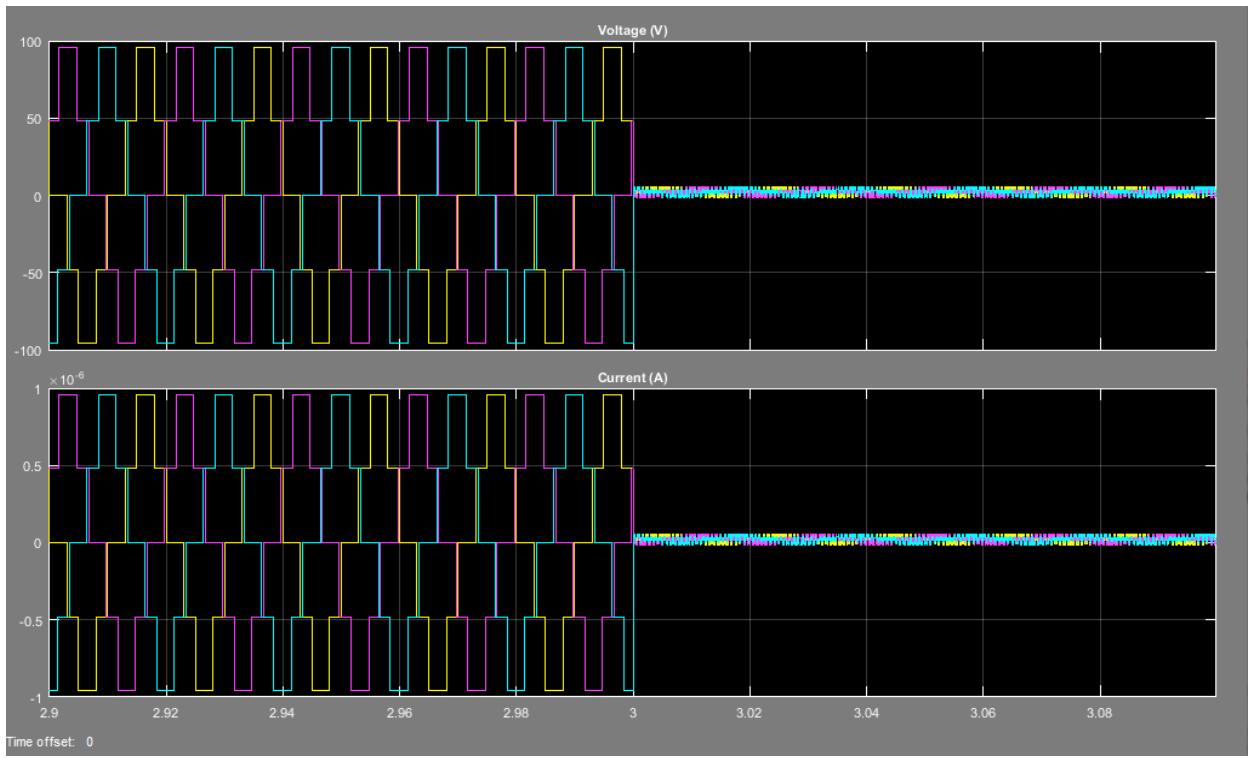

**Fig.19 Output waveform of three-phase 3-level cascaded H-bridge converter among reconfigurable topology 1 with ideal switches with resistance 1e9**

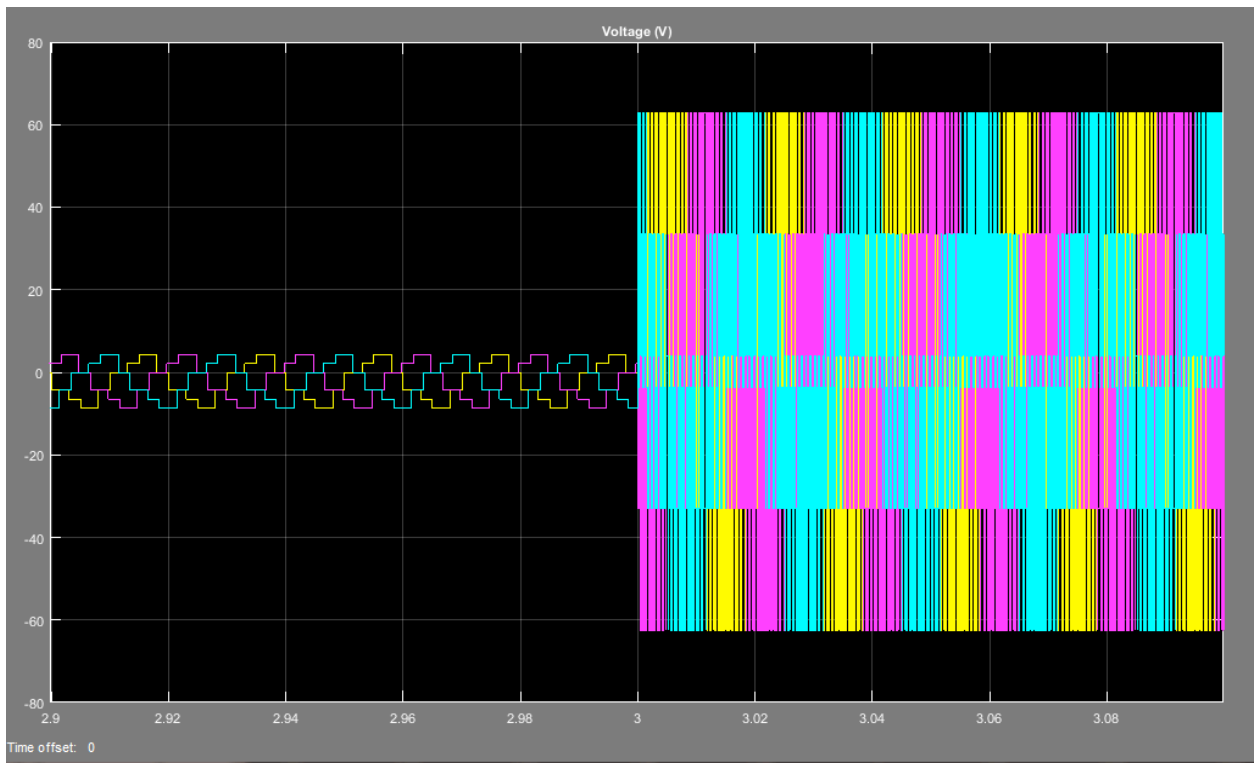

**Fig.20 Output waveform of three-phase 2-level bridge inverter among reconfigurable topology 1 with ideal switches with resistance 1e9**

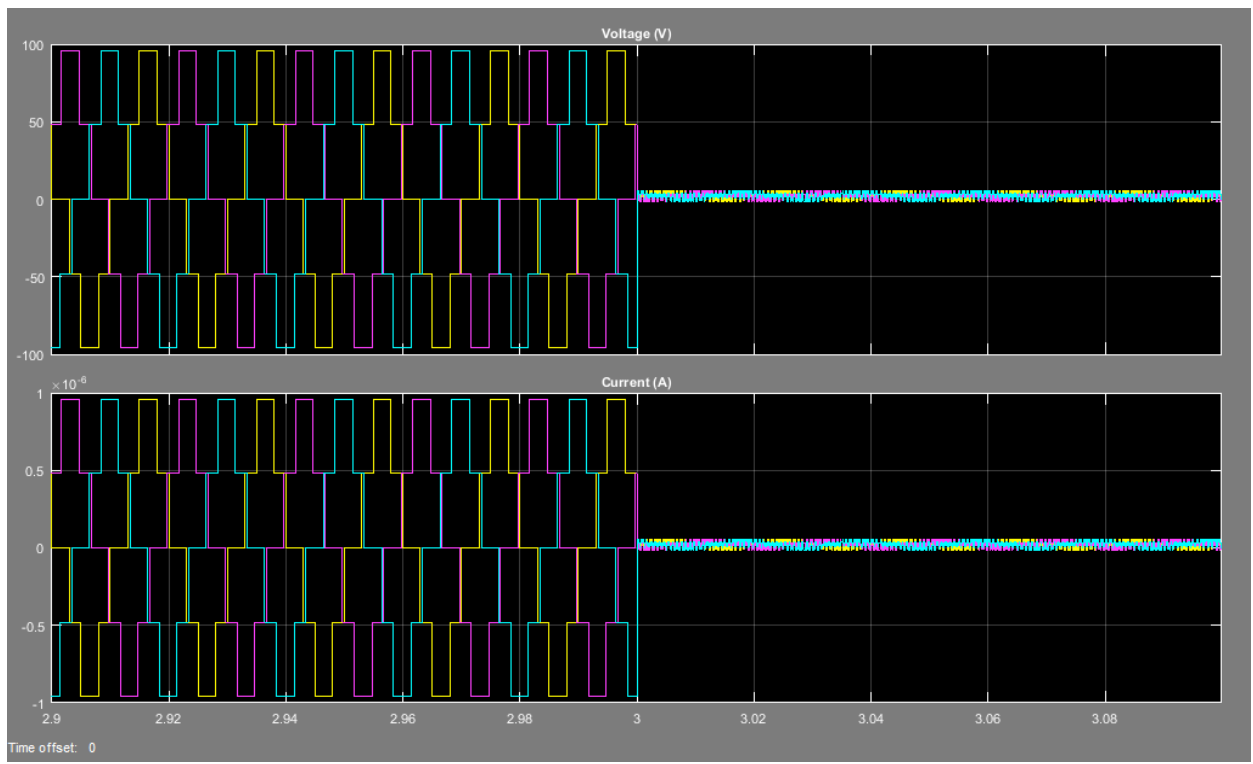

**Fig. 21 Output waveform of three-phase 3-level cascaded H-bridge converter among reconfigurable topology 2 with ideal switches with resistance 1e9**

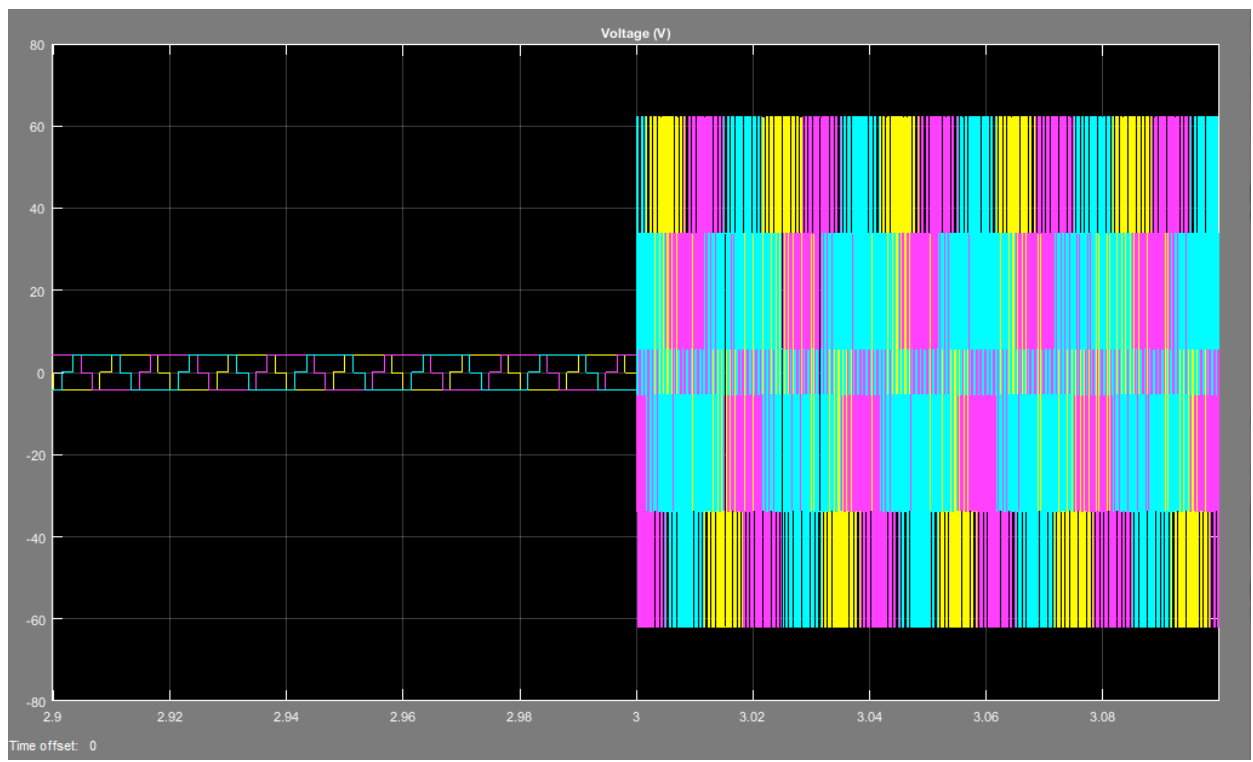

**Fig.22 Output waveform of three-phase 2-level bridge inverter among reconfigurable topology 2 with ideal switches with resistance 1e9**

As the outputs change obviously during the "Off" states and cannot act as ideal as before. It is apparent that the resistances of switches have remarkable influence on the output of the overall circuit and cannot be neglected.

As for the practical simulation time of 10s, reconfigurable topology 1 takes about 42s while it is around 33s for the second topology (simulated by a computer with 8 GB memory and Intel Core i5 processor).

## 4.4 Fault on DC voltage sources

Besides, if one of the upper DC voltage sources is shorted or opened, the 2-level bridge inverter topology can still work normally.

For example, for the reconfigurable topology 1, if the upper DC voltage source of the second parallel line is opened, we can get voltage output waveform as shown in Fig.23 and Fig.24. If the DC voltage source is shorted, similar voltage output waveform will be generated. And the same situation happens to reconfigurable topology 2.

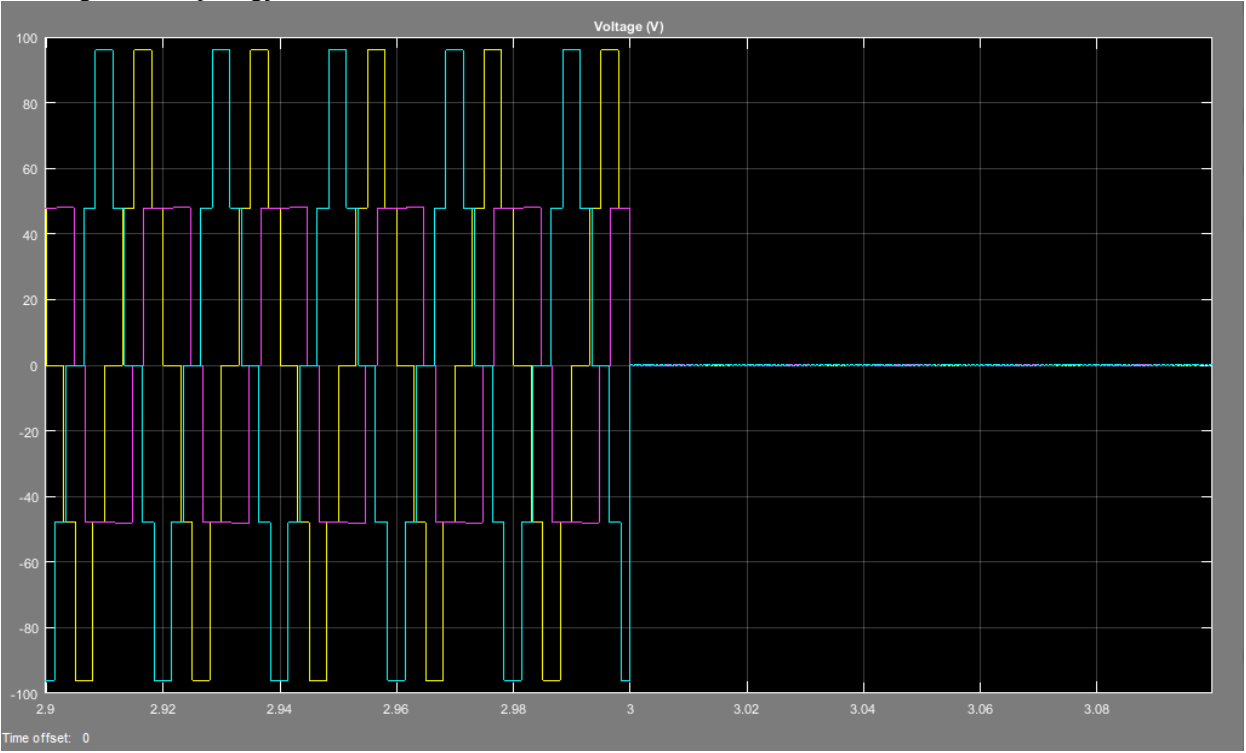

**Fig.23 Output waveform of three-phase 3-level cascaded H-bridge converter among reconfigurable topology 1 with an opened DC voltage source**

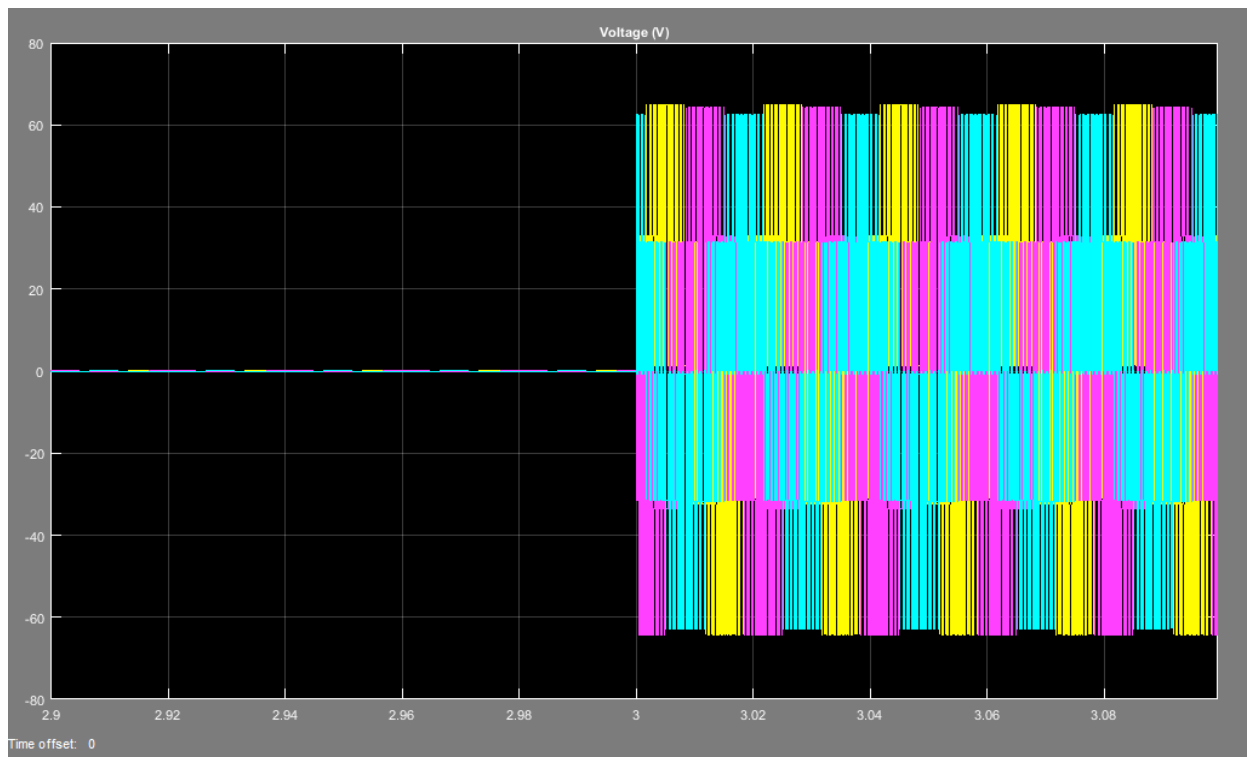

**Fig.24 Output waveform of three-phase 2-level bridge inverter among reconfigurable topology 1 with an opened DC voltage source**

Fig.23 indicates that the phase voltage of the shorted/opened DC voltage source reduces to half of the other two normal phases voltage. And in Fig.24 the waveform is almost the same with all components work normally. Thus, if any DC voltage source breakdown, the reconfigurable power converter can be transformed to the second topology to minimize negative effect on grid.

## 5. Conclusions and Future Work

In this paper, two kinds of reconfigurable power converter topologies are designed, implemented and simulated with MATLAB/SIMULINK. It turned out that they are functional to be utilized in reality. They can be applied to transform between three-phase 3-level cascaded H-bridge converter and three-phase 2-level bridge inverter. This paper provides inspiration and innovation on the design of reconfigurable power electronics topologies.

According to the result of simulation and testing, reconfigurable topology 2 has a more stable and ideal appearance than reconfigurable topology 1. As it is able to work normally when faults occur, such as a voltage source is shorted or opened.

In the future work, the principal and nature about the fault conditions and the further methods to prevent and deal with them are to be researched. Also other further reconfigurable power electronics topologies can be developed based on the ones implemented in this paper.